\documentclass[10pt]{article}

\usepackage{latexsym}
\usepackage{amssymb}
\usepackage{amsfonts}
\usepackage{verbatim}

\def\twi{\~\,$\!$}

\newcommand{\txteqnlabel}{()}
\newenvironment{txteqn}[1]{\renewcommand
{\txteqnlabel}{(#1)}\vspace{1.5ex}{\settowidth
{\parindent}{(#1)}\mbox{}}\hfill\begin {math}\displaystyle}
{\end {math}\hfill \mbox{\txteqnlabel}\vspace {1.5ex}\\}

\newcommand{\bra}[1]{\langle #1 |}
\newcommand{\ket}[1]{| #1 \rangle}
\newcommand{\braket}[2] {\langle #1 | #2 \rangle}

\newcommand{\Tr}{\mathrm{Tr}} 
\newcommand{\bS}{\mathbf S}   
\newcommand{\bP}{\mathbf P}   
\newcommand{\cS}{\mathcal S}  
\newcommand{\cP}{\mathcal P}  
\newcommand{\I} {\mathrm I}   
\newcommand{\amp}{\,\&\,}     
\newcommand{\tensor}{\otimes}

\newcommand {\half} {\frac{1}{2}} 
\newcommand {\quar} {\frac{1}{4}} 

\author {
Alan Macdonald  \\
Department of Mathematics, Luther College \\
Decorah IA, 52101 \ U.S.A.\\ 
macdonal@luther.edu }  

\title{Entanglement, joint measurement, \\and state reduction}

\date{}

\begin {document}
\maketitle
\begin{abstract}
\noindent
Entanglement is perhaps the most important new feature of the quantum world.
It is expressed in quantum theory by the joint measurement formula.
We prove the formula for projection valued observables from a plausible assumption, 
which for spacelike separated measurements 
is a consequence of causality.
State reduction is simply a way to express the joint measurement formula 
after one measurement has been made, and its result known. 
\end{abstract}


\vspace{.05 in}
\normalsize

\newpage
\section{Introduction} \label{sec:intro} 
Entanglement is perhaps the most important new feature of the quantum world.
It is expressed in quantum theory by the joint measurement formula (JMF).
I prove that the JMF is equivalent to the conjunction of two assumptions. 
One is NOEFFECT:
A nonselective measurement of one member of a pair 
of entangled noninteracting systems 
has no effect on measurement probabilities for the other member.
(The measurement is \emph{nonselective} if we do not use its result 
to condition measurement probabilities for the other member.)

For projection valued observables, the JMF is equivalent to NOEFFECT alone.
An example shows that for general observables, NOEFFECT $\nRightarrow$ JMF.

A violation of NOEFFECT in spacelike separated measurements
would allow superluminal communication.
Thus causality implies the JMF 
for spacelike separated measurements of projection valued observables.
The JMF implies violations Bell's inequality, and thus violations of locality.
Thus, within the quantum formalism, \emph{causality implies nonlocality}.

``No signaling'' theorems have eliminated the worry that 
the nonlocality in quantum theory violates causality 
(Jordan, 1983;  Zanchini and Barletta, 1991). 
Our result shows that not only does nonlocality not violate causality, 
it is required to preserve causality.

We prove that the state reduction formula (SRF) 
is an immediate corollary of the JMF:
state reduction is simply a way to express the JMF 
after one measurement has been made, and its result known.
We then prove the von Neumann-L\"uders projection postulate from the SRF.
Thus the ``postulate'' is a theorem, a consequence of the JMF.

All this sheds new light on entanglement, joint measurement, 
state reduction, nonlocality, and causality in quantum theory.

The paper is organized as follows.
Section \ref{sec:qt} reviews the postulates of quantum theory,
without the JMF or SRF. 
Section \ref{sec:JMSR} describes my approach to the JMF and SRF. 
Section \ref{sec:Ozawa} describes Masanao Ozawa's approach to the JMF and SRF
and compares our two approaches.
Section \ref{sec:mp} argues that there is no measurement problem.
Section \ref{sec:example} gives the example 
showing that NOEFFECT $\nRightarrow$ JMF.

\section{QT-}\label{sec:qt}
To prepare for a discussion of the JMF and SRF, 
we review the postulates of quantum theory,
excluding the JMF and SRF.
We call the theory \emph{QT-}.
For more details, see Kraus, 1983 and Busch \emph{et al.}, 1991.

A \emph{quantum system} $\bS$ is represented by a complex Hilbert space 
$\mathsf H_\bS$, which in this paper will be finite dimensional.
A \emph{preparation} of $\bS$ is represented by a \emph{state},
a density operator $\sigma$ on $\mathsf H_\bS$. 
A \emph{measurement} of $\bS$ is represented by an \emph{observable}, 
a positive operator valued measure (POVM) $\cS$. 
Let $\cS$ map the measured value $s$ to $E_s$, 
$0 \le E_s \le \I$.
According to the \emph{measurement formula},
the probability of result $s$ for an $\cS$ measurement on state $\sigma$
is $\Pr(s) = \Tr(E_s\sigma)$.

If $\bS$ is isolated, then $\sigma$ evolves unitarily according 
to \emph{Schr\"odinger's equation}: 
$\sigma \rightarrow U_\bS \,\sigma\, U_\bS^\dagger $.
Important: for now, 
``isolated'' excludes ``entangled with another system''.
The extent to which Schr\"odinger's equation applies to a 
quantum system entangled with another will be the focus of 
\S\ref{sec:Ozawa}. 

Let $\bP$ be another quantum system. 
Then $\bS + \bP$ is represented by 
$\mathsf H_\bS \tensor \mathsf H_\bP$.
Thus the states $\tau$ of $\bS + \bP$ are 
density operators on $\mathsf H_\bS \tensor \mathsf H_\bP$, 
and the observables are POVMs whose values are 
positive operators on $\mathsf H_\bS \tensor \mathsf H_\bP$. 
A measurement of $\cS$ on $\bS + \bP$  
is represented by the POVM which maps $s$ to $E_s \tensor \I$. 
Then from the measurement formula, 
$\Pr(s) = \Tr[(E_s \tensor \I)\tau]$.
The systems $\bS$ and $\bP$ \emph{do not interact} 
if the unitary evolution operator of $\bS + \bP$ factors:
$U_{\bS + \bP} = U_\bS \tensor U_\bP$.

If for some state $\sigma$, 
$\Pr(s) = \Tr(E_s \sigma)$ for every observable $\cS$ and every result $s$, 
then $\sigma$ is the state of $\bS$.
For the $\Tr(E_s \sigma)$ uniquely determine 
the state $\sigma$.
We say that ``probabilities determine states''.

For reference we list several identities which we will use 
without comment:
$\Tr(XY) = \Tr(YX)$,
$\braket{s_1 \tensor p_1}{s_2 \tensor p_2} 
           = \braket{s_1}{p_1} \braket{s_2}{p_2}$,
$X \tensor Y = (X \tensor \I)(\I \tensor Y)$, and
$(X \tensor Y)\ket{s \tensor p} = X \ket{s} \tensor Y \ket{p}$.
The \emph{partial trace} operator $\Tr_\bP$ maps operators on 
$\bS + \bP$ to operators on $\bS$ 
(Cohen-Tannoudji \emph{et al.}, 1997). 
We have the partial trace identities 
$\Tr(X) = \Tr[\Tr_{\bP}(X)]$ and
$\Tr_{\bP}[(X \tensor \I)Y] = X\Tr_{\bP}(Y)$ 
(Kraus, 1983; Busch \emph{et al.}, 1991).
Using these identities and ``probabilities determine states'', we see that 
if the state of $\bS + \bP$ is $\tau$, 
then the state of $\bS$ is $\Tr_{\bP}(\tau)$:
\begin{equation}  
\Pr(s) = \Tr[(E_s \tensor \I)\tau] 
       = \Tr\left\{\Tr_{\bP}\left[(E_s \tensor \I)\tau\right]\right\} 
       = \Tr\left[E_s\, \Tr_{\bP}(\tau)\right].     \label{eq:stateS}
\end{equation}

\section{Joint measurement and state reduction} \label{sec:JMSR}

In this section we prove results about joint measurement, state reduction, 
causality, and nonlocality in the theory QT- defined in \S\ref{sec:qt}.

\vspace{.1 in}
\textbf{Joint Measurement Formula.} 
\emph{Prepare $\bS + \bP$ in state $\tau$ at time $t_1$,
after which $\bS$ and $\bP$ do not interact.
At time $t_\bP \geq t_1$ measure observable $\cP$ of $\bP$, with result $p$. 
At time $t_\bS \geq t_1$ measure observable $\cS$ of $\bS$, with result $s$.
Let $U_\bP$ be the unitary evolution operator for $\bP$ from $t_1$ to $t_\bP$.
Let $U_\bS$ be the unitary evolution operator for $\bS$ from $t_1$ to $t_\bS$. 
Then }

\begin{txteqn}{JMF}
\Pr(s \amp p) =
  \Tr\left[\left(U_{\bS}^\dagger E_sU_{\bS} \tensor 
                 U_{\bP}^\dagger E_p U_{\bP}  \right)\tau \right].
\end{txteqn}

For given $t_\bP, \cP, t_\bS, \mathrm{and\ } \cS$
let the POVM representing the joint measurement map the result 
$(s, p)$ to $E_{s \amp p}$. 
Then according to the measurement formula,
$\Pr(s \amp p) = \Tr(E_{s \amp p}\tau)$ for all $s, p$, and $\tau$. 
Thus the JMF for the measurement is equivalent to
\begin{equation}  
\forall s,p \ \ E_{s \amp p} =
    U_{\bS}^\dagger E_sU_{\bS} \tensor U_{\bP}^\dagger E_p U_{\bP}.
\label{eq:JMF}
\end{equation}
The (nonselective) probability of $s$ is 
$\sum_p\Pr(s \amp p) = \Tr \left[ \left(\sum_pE_{s \amp p}\right) \tau \right]$.
If the $\cP$ measurement is not made, then according to Eq. (\ref{eq:sch}), 
the probability of $s$ is 
$$\Tr\left[E_s \left(U_\bS \Tr_\bP(\tau)U_\bS^\dagger \right) \right] = 
\Tr\left[ \left(U_\bS^\dagger E_s U_\bS \tensor \I \right) \tau \right].$$
NOEFFECT from \S\ref{sec:intro} asserts that the two probabilities are equal:
\begin{quote}
A nonselective measurement of one member of a pair 
of entangled noninteracting systems 
has no effect on measurement probabilities for the other member.
\end{quote}
Thus according to NOEFFECT,

\begin{txteqn}{NOEFFECT}
\forall s \ \ \sum_p E_{s \amp p} = U_\bS^\dagger E_s U_\bS \tensor \I.
\end{txteqn}
Similarly,

\begin{txteqn}{NOEFFECT}
\forall p \ \ \sum_s E_{s \amp p} = \I \tensor U_\bP^\dagger E_p U_\bP.
\end{txteqn}

Consider also the assertion that 
$E_{s \amp p}$ is the product of its marginals:

\begin{txteqn}{PRODMARG}
\forall s,p \ \ \ E_{s \amp p} = \left( \sum_{p} E_{s \amp p} \right) 
               \left( \sum_{s} E_{s \amp p} \right).
\end{txteqn}
\vspace{.1 in}

\textbf{Theorem 1}.
 \emph{For given $t_\bP, \cP, t_\bS$, and $\cS$},
 $$\mathrm{JMF} \Leftrightarrow \mathrm{(NOEFFECT + PRODMARG).}$$

\vspace{.05 in}
\emph{Proof.} We use the JMF in the form Eq. (\ref{eq:JMF}).

JMF $\Rightarrow$  NOEFFECT. 
Sum Eq. (\ref{eq:JMF}) over $p$ and use $\sum_pE_p = \I$.
(This is the no signaling theorem of Jordan, 1983.)

JMF $\Rightarrow$ PRODMARG. 
Multiply the two NOEFFECT equations,
which we have just shown follow from the JMF, 
and use Eq. (\ref{eq:JMF}) to obtain PRODMARG.

(NOEFFECT + PRODMARG) $\Rightarrow$ JMF. Multiply the two NOEFFECT equations 
and use PRODMARG to obtain Eq. (\ref{eq:JMF}).  $\ \ \ \ \Box$

\vspace{.1 in}
\textbf{Corollary 2.}  \emph{If $\cP$ and $\cS$ are projection valued,
then} JMF $\Leftrightarrow$ NOEFFECT.

\vspace{.05 in}
\emph{Proof.} 
From the theorem, it is sufficient to prove that 
if $\cP$ and $\cS$ are projection valued, then NOEFFECT $\Rightarrow$ PRODMARG.
For a projection valued $\cS$, the $E_s$ are orthogonal projections.
Thus the $U_\bS^\dagger E_s U_\bS \tensor \I$ 
on the right side of the first NOEFFECT equation are orthogonal projections. 
Sums of these projections are projections.
Every POVM on a product space with projection valued marginal measures 
satisfies PRODMARG (Davies, 1976, Th. 2.1, Eq. 2.7).  $\ \ \ \Box$
 
\vspace{.1 in}
The example $E^{\,\prime}$ of \S\ref{sec:example} shows that 
for general POVMs, NOEFFECT $\nRightarrow$ JMF.

The implication NOEFFECT $\Rightarrow$ JMF for projection valued observables
is of special interest. 
As noted in \S\ref{sec:intro}, for spacelike separated measurements 
causality implies NOEFFECT.
Thus, in QT-: 

\vspace{.1 in}
\textbf{Corollary 3}. \emph{Causality implies the JMF
for spacelike separated measurements of projection valued observables}.

\vspace{.05 in}
The JMF predicts violations of Bell's inequality 
for some spacelike separated measurements of projection valued observables.
It thus predicts violations of locality.
Thus, in QT-:

\vspace{.1 in}
\textbf{Corollary 4}. \emph{Causality implies nonlocality}.

\vspace{.05 in}
We now turn to the SRF. Since probabilities determine states,
we can reformulate NOEFFECT:
\begin{quote}
A nonselective measurement of one member of a pair 
of entangled noninteracting systems 
has no effect on the state of the other member.
\end{quote}
But the SRF says that if we make a \emph{selective} measurement, 
conditioning the state of $\bS$ on the $\cP$ measurement result, 
then we must \emph{reduce} the state of $\bS$:

\vspace{.1 in}
\textbf{State Reduction Formula.} 
\emph{Prepare $\bS + \bP$ in state $\tau$ at time $t_1$,
after which $\bS$ and $\bP$ do not interact.
At $t_1$ measure observable $\cP$ of $\bP$, with result $p$. 
Let $U_\bS$ be the unitary evolution operator of $\bS$ 
over the time of the $\cP$ measurement. 
Let $\sigma_p$ be the state of $\bS$ after the $\cP$ measurement,
conditioned on $p$. Then}

\begin{txteqn}{SRF}
  \sigma_p = U_{\bS} \frac
    {\Tr_\bP\!\left[(\I \tensor E_p)\tau\right]}
    {\Tr      \left[(\I \tensor E_p)\tau\right]} U_{\bS}^\dagger.
\end{txteqn}
\textbf{Remarks.} 
(i) The SRF requires \emph{no} assumptions 
about the state of $\bP$ after the $\cP$ measurement, 
even that $\bP$ still exists.
(ii) Since we do not assume that Schr\"odinger's equation applies 
to a system entangled with another, 
we cannot interpret the SRF as giving the evolution 
of $\bS$ during the $\cP$ measurement.
(iii) It is \emph{classical} information, i.e., $p$, 
which allows us to reduce the state of $\bS$ to $\sigma_p$.
(iv) From the SRF,
$\sum_p \Pr(p) \sigma_p = U_{\bS} \Tr_\bP(\tau) U_{\bS}^\dagger$,
the unreduced state.

\vspace{.1 in}
\textbf{Theorem 5.} JMF $\Rightarrow$ SRF.

\vspace{.05 in}
\emph{Proof.}
Measure $\cS$ immediately after the $\cP$ measurement. From the JMF,
\begin{equation}  
\Pr(s \amp p) = \Tr\,\{(U_{\bS}^\dagger E_sU_{\bS} \tensor E_p)\tau\}.
\label{eq:JMFSR}
\end{equation}
Thus for every $\cS$ and every $s$,
\begin{eqnarray}  
     \Pr\left(s \,|\, p \right)
 &=& \frac{\Pr\left(s \amp p \right)}{\Pr\left(p\right)} 
  =  \frac{\Tr\left\{\left( U_\bS^\dagger E_s U_\bS \tensor E_p \right)
                                                         \tau \right\}}
    {\Tr\,\left[ \left( \I \tensor E_p \right) \tau \right]} 
	                                                       \nonumber \\  
&=& \frac{\Tr\left\{\Tr_\bP \left[\left( U_\bS^\dagger E_s U_\bS \tensor \I \right) 
                  \left( \I \tensor E_p \right) \tau \right]\right\}}
    {\Tr\,\left[ \left( \I \tensor E_p \right) \tau \right]} \label{eq:ess2p}  \\  
&=& \Tr\left\{ E_s \left( \,U_{\bS} \frac
    {\Tr_\bP\!\left[(\I \tensor E_p)\tau\right]}
    {\Tr      \left[(\I \tensor E_p)\tau\right]} 
     U_{\bS}^\dagger \right) \right\}.\ \ \ \ \ \nonumber 
\end{eqnarray}
Since probabilities determine states, the SRF follows. $\ \ \ \Box$

\noindent
(For more on this kind of reasoning to obtain state reduction, 
see Svetlichny, 2002.)

Conversely, given the SRF, a rearrangement of
Eq. (\ref{eq:ess2p}) proves Eq. (\ref{eq:JMFSR}).
Thus 
\begin{quote} 
\emph{State reduction is simply a way to express the JMF 
after one measurement has been made, and its result known.}
\end{quote}

K. Kraus makes a similar statement: ``[State reductions] provide a convenient 
`shorthand' description of correlation measurements. 
We may thus conclude that, contrary to widespread belief, 
[state reductions] can be perfectly well understood,
\emph{if} quantum mechanics is assumed 
to be valid also for measuring instruments.'' (Kraus, 1983, p. 99; my emphasis.) 
Our proof of the SRF does not assume that quantum mechanics 
is valid for measuring instruments.
Thus Kraus' \emph{if} clause is unnecessary.
For more on this, see \S\ref{sec:mp}.

\vspace{.1 in}
\textbf{Corollary 6.} \emph{If $\cP$ is projection valued, then}
NOEFFECT $\Rightarrow$ SRF.

\vspace{.05 in}
\emph{Proof.} 
Measure a projection valued observable $\cS$ immediately after the $\cP$ measurement.
Then Corollary 2 implies Eq. (\ref{eq:JMFSR}),
which implies Eq. (\ref{eq:ess2p}) for projections $E_s$,
which is sufficient to imply the SRF for the $\cP$ measurement. 
$\ \ \Box$

\vspace{.1 in}
We close this section with a discussion of 
the von Neumann-L\"uders measurement model.
Let $\bS$ be a quantum \emph{system} to be measured 
and $\bP$ be a quantum \emph{probe}, 
which is part of a macroscopic measuring apparatus. 
Initially $\bS$ and $\bP$ are separated and unentangled, 
and in states $\sigma_0$ and $\pi_0$. 
The system enters the measuring apparatus, interacts with the probe, 
and leaves the apparatus.
Let $\tau = U(\sigma_0\, \tensor \pi_0)\,U^\dag$ 
be the state of $\bS + \bP$ after the interaction,
which is called a \emph{premeasurement}.
(A premeasurement is \emph{not} a measurement:
a premeasurement is reversible and no measured value is created.)
Now measure $\cP$, with the result $p$ appearing on the measuring apparatus.
In the von Neumann-L\"uders model, 
the $\cP$ measurement serves as a proxy for an $\cS$ measurement.

The model is for projection valued $\cS$ 
with an associated self-adjoint operator
$\sum_{ij} s_i \ket{s_{ij}} \bra{s_{ij}}$. 
Let $\cP$ be a nondegenerate projection valued observable 
with an associated self-adjoint operator $\sum_i p_i \ket{p_i} \bra{p_i}$.
Choose a unitary operator $U$ with $U(\ket{s_{ij}} \ket{p_0}) = \ket{s_{ij}} \ket{p_i}$
for some fixed initial state $\ket{p_0}$ of $\bP$.
Then for an initial vector state 
$\ket{s_0} = \sum_{ij} a_{ij} \ket{s_{ij}}$ of $\bS$,
$U(\ket{s_0} \ket{p_0}) = \sum_{ij} a_{ij} \ket{s_{ij}} \ket{p_i} \equiv \ket{t}$.
For a  $\cP$ measurement on state $\ket{t}$,   $\Pr(p_k) = \sum_j|a_{kj}|^2$. 
For an $\cS$ measurement on state $\ket{s_0}$, $\Pr(s_k)$ has the same value.
Thus a  $\cP$ measurement on state $\ket{t}$   with result $p_k$ is 
also an $\cS$ measurement on state $\ket{s_0}$ with result $s_k$.

The SRF gives the reduced state $\sigma_{s_k}$ of $\bS$ 
after the $\cS$ measurement.
To apply it, we first use the identity 
$\Tr_\bP[(\I \tensor X)Y] = \Tr_\bP[Y(\I \tensor X)]$ (Kraus, 1983, Eq. 5.15):
\begin{eqnarray*}  
      \Tr_\bP \left\{ \left( \I \tensor E_{p_k}  \right)\tau \right\}
  &=& \Tr_\bP \left\{ \left( \I \tensor E_{p_k}  \right) \!
          \ket{t} \, \bra{t}(\I \tensor E_{p_k}) \right\}  \\
  &=& \Tr_\bP \left\{ \Sigma_j a_{kj}\ket{s_{kj}} \ket{p_k} \; 
                \Sigma_j \bar{a}_{kj}\bra{s_{kj}} \bra{p_k}  \right\}  \\
  &=&  \Tr_\bP\left\{ E_{s_k}\ket{s_0} \ket{p_k} \,
	                 \bra{s_0} E_{s_k}  \bra{p_k} \right\}  \\
  &=&   E_{s_k}\ket{s_0} \bra{s_0}  E_{s_k} .     
\end{eqnarray*}
Substitute this into the SRF:
$$\sigma_{s_k} =
          U_{\bS}\, \frac{ E_{s_k} \ket{s_0} \, \bra{s_0}E_{s_k} }
               {\Tr\left\{ E_{s_k} \ket{s_0} \, \bra{s_0}E_{s_k} \right\}}
                                                          \, U_{\bS}^\dagger. $$
As a vector, the reduced state is 
$U_{\bS} E_{s_k} \ket{s_0}/\|E_{s_k} \ket{s_0}\|$.
This is the state given by the \emph{von Neumann-L\"uders projection postulate}.
Since JMF $\Rightarrow$ SRF, the ``postulate'' is a \emph{theorem} of QT- + JMF.

\section{Ozawa's approach}\label{sec:Ozawa} 
Masanao Ozawa has published several papers on joint measurement and state reduction 
(Ozawa, 1997a, 1997b, 1998a, 1998b, 2000a, 2000b, 2000c). 
He argues, correctly I believe, that existing proofs of the JMF 
and SRF are inadequate or flawed 
(Ozawa, 2000a, p. 6; 1998a, p. 616; 1997b, p. 123; 1997a, p. 233). 
He then offers his own proofs of the JMF 
(Ozawa, 1997a, Th. 5.1; 2000a, Th. 3)
and the SRF
(Ozawa 1998a, Eq. 32; 1997b, Eq. 43).
Ozawa considers projection valued observables only. 

As emphasized in \S\ref{sec:qt}, QT- does not assume that 
Schr\"odinger's equation applies to a quantum system entangled with another.
But we can prove: 
\begin{quote}
A unitary evolution of one member of a pair of entangled noninteracting 
systems has no effect on the state of the other member.
\end{quote}
\emph{Proof.} Since $\bS$ and $\bP$ do not interact, 
the unitary evolution operator of $\bS + \bP$ factors: 
$V_{\bS + \bP} = V_\bS \tensor V_\bP$. 
Let $\tau$ be the initial state of $\bS + \bP$.
Then for all $E_s$,
\begin{eqnarray}  
     \Tr[ E_s(V_\bS \Tr_\bP(\tau) V_\bS ^\dagger)]  
 &=& \Tr[(E_s \tensor \I)        (V_\bS \tensor \I) \tau 
                                 (V_\bS ^\dagger \tensor \I)]    
										                              \nonumber\\
 &=& \Tr[(E_s \tensor \I)(\I \tensor V_\bP^\dagger) (\I \tensor V_\bP)
       (V_\bS \tensor \I) \tau   (V_\bS ^\dagger \tensor \I)]\nonumber\\
 &=& \Tr[(E_s \tensor \I)        (V_\bS \tensor V_\bP) \tau 
                                 (V_\bS \tensor V_\bP)^\dagger]
                                    \label{eq:unitary}      \label{eq:sch}\\
 &=& \Tr\{E_s \Tr_\bP[           (V_\bS \tensor V_\bP) \tau 
                                 (V_\bS \tensor V_\bP)^\dagger]\}. \nonumber
\end{eqnarray}
Thus the state of $\bS$ at a later time, 
$\Tr_\bP [(V_\bS \tensor V_\bP) \tau (V_\bS \tensor V_\bP)^\dagger]$, 
is the same as the state given by Schr\"odinger's equation 
applied to $\bS$ alone, $V_\bS \Tr_\bP(\tau) V_\bS ^\dagger$.
$\ \ \ \Box$ \\
(This is the no signaling theorem of Zanchini and Barletta, 1991, Th. 3.)

For projection valued observables, we proved the JMF in Corollary 2 and 
the SRF in Corollary 6 from the assumption NOEFFECT:
\begin{quote} 
A nonselective measurement of one member of a pair 
of entangled noninteracting systems 
has no effect on the unreduced state of the other member.
\end{quote} 
(We use the reformulated version following Corollary 4 
and add the word ``unreduced'' for clarity and comparison.)

Ozawa uses a different assumption:
\begin{quote}
A selective measurement of one member of a pair 
of entangled noninteracting systems has no effect 
on the reduced state of the other member.
\end{quote}
(In Ozawa, 1998a see the discussions surrounding Eqs. (5), (6), and (15), 
and also p. 622.)

One example of Ozawa's use of his assumption is in his proof of the JMF
in Ozawa, 1997a, Th. 5.1, when passing from the third to the fourth
member in the equation between Eqs. (9) and (10).
(Ozawa has confirmed this reading in a private communication.)
Another example is in his proof of the SRF in Ozawa, 1998a, Sec. 7.

Ozawa agrees that the SRF gives the reduced state $\sigma_p$ 
after the $\cP$ measurement,
but his assumption rules out our view that the reduction 
occurs with the measurement,
a view he rejects (Ozawa, 1997b, p. 123).
For him, the reduction occurs earlier, with the \emph{premeasurement},
to a state which we denote $\sigma_p^1$.
($\sigma_p^1$ is denoted 
$\rho(t + \Delta t \,|\, \mathbf{a}(t) \in \{p\})$ in Ozawa, 2000a,
and $\rho(t + \Delta t \,|\, p)$ in Ozawa, 1998a and 1997b.)
(Warning: Ozawa sometimes calls just the premeasurement 
-- which he calls \emph{stage 1} -- a ``measurement'' 
(Ozawa, 1998a, Eq. (1); 1997b, Eq. (1))).

According to Ozawa, $\sigma_p^1$ is the state of $\bS$ 
after the premeasurement, 
``conditional upon'' the result $p$ of the later
$\cP$ measurement (Ozawa, 2000a, p. 9), 
or ``that leads to the outcome $p$'' in the measurement (Ozawa, 1997b, p. 124).
More specifically:
\begin{quote} \small
Suppose the system and probe are spin-$\frac{1}{2}$ 
particles brought into the singlet state by the premeasurement. 
After the premeasurement is complete, we can choose to measure  
the spin of the probe in the $z$-direction or the $x$-direction.  
If we choose the $z$-direction and the result is ``up'', 
then the system was prepared in the ``down'' eigenstate 
$\sigma_{\!\, \downarrow}^1$ just after the premeasurement.  
If we choose the $x$-direction and the result is ``left'', 
then the system was prepared in the ``right'' eigenstate 
$\sigma_{\, \rightarrow}^1$ just after the premeasurement.
[Private communication.]
\end{quote} \normalsize

If, according to Ozawa's assumption, $\bS$ evolves unitarily from 
after the premeasurement until after the probe measurement, 
and if its state after the probe measurement is $\sigma_p$, 
then its state after the premeasurement is, from the SRF,
$$\frac{\Tr_\bP\!\left[(\I \tensor E_p)\tau\right]}
       {\Tr      \left[(\I \tensor E_p)\tau\right]}. $$
This is Ozawa's expression for $\sigma_p^1$ 
(Ozawa, 1998a, Eq. (32); 1997b, Eq. (34)). 
For him, the SRF describes a unitary
evolution of $\bS$ from $\sigma_p^1$ to $\sigma_p$. 
For me, the SRF does not describe an evolution of $\bS$,
as stated in the remarks following the SRF.

Bell's inequality is relevant here.
The inequality shows that not only is the result $p$ of the probe measurement 
not known before the measurement, \emph{it does not exist} before the measurement.
This even though $p$ would be correlated with the result of a 
later measurement of $\cS$. 
Mermin explains this clearly (Mermin, 1981 and 1985).

For me, this makes the states $\sigma_p^1$ problematic.
Furthermore, they are not needed to obtain the SRF: 
we proved in \S\ref{sec:JMSR} that the correlations given by the JMF 
imply that the state of $\bS$ after the $\cP$ measurement is given by the SRF. 
State reduction is not a \emph{dynamical} consequence of Schr\"odinger's equation; 
it is a \emph{logical} consequence of entanglement.

To reject attributing the state reduction of $\bS$ to the $\cP$ measurement
is to cling to classical notions of causality,
instead of fully embracing that remarkable new quantum phenomenon, 
entanglement.

\section{The measurement problem}\label{sec:mp}

We have been careful to distinguish the probe $\bP$, 
a \emph{quantum} system,
from the macroscopic apparatus measuring it.
We made no assumptions about the apparatus other than the minimal requirement 
that it display measurement results in accordance with the measurement formula.
In particular, we did not model it as a 
quantum system obeying Schr\"odinger's equation.
Modeling the apparatus in this way leads to the notorious 
\emph{measurement problem}: the appearance of a definite measured value 
on the apparatus would be a state reduction of the apparatus, 
which is inconsistent with Schr\"odinger's equation.

I argue at length elsewhere that the apparatus cannot be so modeled
and thus there is no measurement problem (Macdonald, 2002).
Here I support this point of view only with the following quotes. 

In \emph{The Quantum Theory of Measurement}, 
P. Busch, P. Lahti, and P. Mittelstaedt write: 
``The quantum theory of measurement is motivated by the idea of the 
universal validity of quantum mechanics, according to which this theory 
should be applicable, in particular, to the measuring process. 
One would expect, and most researchers in the foundations of quantum mechanics 
have done so, that the problem of measurement should be solvable 
\emph{within} quantum mechanics. 
The long history of this problem shows that ... 
there seems to be no straightforward route to its solution.''  
(Busch \emph{et al.}, 1991, p. 138)  

K. Kraus also describes the measuring apparatus as a quantum system 
(Kraus, 1983, pp. 81, 99). But ``There are good reasons to doubt that 
quantum mechanics in its present form is the appropriate theory of 
macroscopic systems.'' (Kraus, 1983, p. 100)

According to A. Leggett, 
``What is required is to explain how one particular macrostate 
can be forced by the quantum formalism to be realized. 
In the opinion of the present author 
(which is shared by a small but growing minority of physicists) 
no solution to this problem is possible within the framework of 
conventional quantum mechanics.'' (Leggett, 1992, p. 231)

W. Zurek writes, 
``The key (and uncontroversial) fact has been known almost since 
the inception of quantum theory, but its 
significance ... is being recognized only now: 
macroscopic systems are never isolated from their environment. 
Therefore they should not be expected to follow 
Schr\"odinger's equation, which is applicable only to a closed system.'' 
(Zurek, 1991)

\section{NOEFFECT $\nRightarrow$ PRODMARG} 
\label{sec:example} 

Consider the following measurement.
A spin-$\half$ particle $\bS$ moving in the $y$-direction 
enters a Stern-Gerlach device oriented in the $z$-direction.
In each output beam ($\pm z$) there is a SG device 
oriented in the $x$-direction. 
Detect $\bS$ leaving one of the $x$-direction SG devices.
Assign a value 0 to the measurement if $\bS$ 
is detected in a $-x$ beam and a 1 if in a $+x$ beam. 
Then for every state of $\bS$, $\Pr(0) = \Pr(1) = \half$.
Think of this triple SG device as a fair coin tosser.
The POVM $E_0 = E_1 = \half \I$ represents the measurement: 
for every state $\sigma$ of $\bS$, $\Tr(E_0 \sigma) = \Tr(E_1 \sigma) = \half$.

Let $\bP$ be another spin-$\half$.
Measure both $\bS$ and $\bP$ with triple SG devices. 
Absent any assumption about the joint measurement probabilities,
we can imagine different POVMs giving those probabilities.
One possibility is $E_{s \amp p}$ with 
$E_{0 \amp 0} = E_{0 \amp 1} = E_{1 \amp 0} = 
E_{1 \amp 1} = \quar \I \tensor \I$. 
Another is $E^{\,\prime}_{s \amp p}$ with 
$E^{\,\prime}_{0 \amp 0} = E^{\,\prime}_{1 \amp 1} = \half \I \tensor \I$ and 
$E^{\,\prime}_{0 \amp 1} = E^{\,\prime}_{1 \amp 0} = 0$. 
For every state of $\bS + \bP$, 
$E_{s \amp p}$ predicts two \emph{independent} fair coin tosses
and $E^{\,\prime}_{s \amp p}$ predicts two \emph{correlated} fair coin tosses, 
0 with 0 and 1 with 1. 

Straightforward calculations show that $E_{s \amp p}$ satisfies 
both NOEFFECT and PRODMARG.
From these, we can see that the JMF implies that $E_{s \amp p}$ 
represents the joint measurement:
\begin{eqnarray*}  
\Pr(s \amp p) &=& \Tr\left[\left( E_s \tensor E_p \right)\tau \right] 
 =  \Tr\left[\left( E_s \tensor \I \right)\left(\I \tensor E_p \right)\tau \right]\\
&=& \Tr\left[\left( \sum_p E_{s \amp p} \right)
             \left(\sum_s E_{s \amp p}\right)\tau \right]
 = \Tr ( E_{s \amp p} \tau ).
\end{eqnarray*}

The POVM $E^{\,\prime}_{s \amp p}$ satisfies NOEFFECT but not PRODMARG.
Thus NOEFFECT $\nRightarrow$ PRODMARG.

\vspace{.05 in}
\textbf{Acknowledgments.} 
I thank Professor Masanao Ozawa for a lengthy and helpful correspondence. 
I also thank Martin Barrett and Normann Plass for helpful comments.

\vspace{.2 in}
\Large{\textbf{References.}}
\normalsize

\vspace{.1 in}
\small
Busch, P., Lahti, P., and Mittelstaedt P. (1991).
\emph{The quantum theory of measurement\/}, 
Springer-Verlag, Berlin.

Cohen-Tannoudji, C., Diu, B., and Lalo\"e, F. (1977). 
\emph{Quantum Mechanics}, 
Herman/Wiley, Paris.

Davies,  E. B. (1976).
\emph{Quantum theory of open systems\/}, 
Academic Press, London.

Jordan, T. (1983). 
Quantum correlations do not transmit signals, 
Phys. Lett. A \textbf{94}, 264.

Kraus, K. (1983).
\emph{States, effects, and operations\/}, 
Springer-Verlag, Berlin.

Leggett, A. J. (1992).  
On the Nature of Research in Condensed-State Physics,    
\emph{Found. Phys.\/} \textbf{22}, 221.

Macdonald, A. (2002)
Quantum theory without measurement or state reduction problems, 
http://faculty.luther.edu/\twi macdonal.

Mermin, D. (1981).
Bringing home the atomic world, 
\emph{Am. J. Phys.\/} \textbf{49}, 940.

Mermin, D. (1985).
Is the moon there when nobody looks?, 
\emph{Phys. Today\/}, \textbf{38} (4), 38. 

Ozawa, M. (1997a).
Quantum State Reduction and the Quantum Bayes Principle, in: 
O. Hirota \emph{et al.}, eds., 
\emph{Quantum Communication, Computing and Measurement\/}, 
Plenum, New York. Also quant-ph/9705030. 

Ozawa, M. (1997b).
An Operational Approach to Quantum State Reduction, 
\emph{Ann. Phys.\/} \textbf{259}, 121. 
Also quant-ph/9706027.

Ozawa, M. (1998a).
Quantum State Reduction: An Operational Approach, 
\emph{Fortschr. Phys.\/} \textbf{46}, 615. 
Also quant-ph/9711006.  

Ozawa, M. (1998b).
On the Concept of Quantum State Reduction: Inconsistency of the Orthodox View, 
quant-ph/9802022. 

Ozawa, M. (2000a).
Operational characterization of simultaneous measurements in quantum mechanics, 
\emph{Phys. Lett. A\/} \textbf{275}, 5. 
This is a letter version of Ozawa, 2000c. 
A similar paper is available as quant-ph/9802039.

Ozawa, M. (2000b).
Measurements of nondegenerate discrete observables, 
\emph{Phys. Rev. A\/} \textbf{62}, 062101. 
Also quant-ph/0003033.

Ozawa, M. (2000c).
Operations, Disturbance, and Simultaneous Measurability, 
quant-ph/0005054. 

Svetlichny, G. (2002).
Causality implies formal state collapse,
quant-ph/0207180.

Zanchini, E. and Barletta, A. (1991).
Absence of Instantaneous Transmission of Signals in Quantum Theory of Measurement, 
\emph{N. Cimento B\/} \textbf{106}, 419.

Zurek, W. (1991). 
Decoherence and the transition from quantum to classical, 
\emph{Phys. Today\/} \textbf{44} (10), 36.

\end{document}